# Investigating Deep Neural Network Architecture and Feature Extraction Designs for Sensor-based Human Activity Recognition

Danial Ahangarani
Department of Computer Engineering,
Sharif University of Technology
Email: Daniahangarani@sharif.edu

Mohammad Shirazi
Department of Statistics,
Mathematics and Computer,
Allameh Tabataba'i University
Email: shirazi_m@atu.ac.ir

Navid Ashraf
Department of Statistics,
Mathematics and Computer,
Allameh Tabataba'i University
Email: navid_ashraf@atu.ac.ir

*Abstract*— **The extensive ubiquitous availability of sensors in smart devices and the Internet of Things (IoT) has opened up the possibilities for implementing sensor-based activity recognition. As opposed to traditional sensor time-series processing and hand-engineered feature extraction, in light of deep learning's proven effectiveness across various domains, numerous deep methods have been explored to tackle the challenges in activity recognition, outperforming the traditional signal processing and traditional machine learning approaches. In this work, by performing extensive experimental studies on two human activity recognition datasets, we investigate the performance of common deep learning and machine learning approaches as well as different training mechanisms (such as contrastive learning), and various feature representations extracted from the sensor time-series data and measure their effectiveness for the human activity recognition task.**

*Keywords: human activity recognition; deep learning; contrastive learning; sensors; pretraining*

## I. Introduction

The recent advancements in human activity recognition have given rise to a wide range of applications, which include smart homes [1], efficient manufacturing environments [4,5], and patient activity monitoring for healthcare applications [3]. Activity recognition plays a crucial role in human life by capturing people's behaviors through data, enabling computing systems to monitor, analyze, and assist them in their daily activities. Due to the availability of various sensors such as accelerometers and gyroscopes (i.e., inertial measurement units or IMUs) in most off-the-shelf smart devices, as opposed to video-based approaches [6], recent approaches for human activity recognition have relied on such sensors [7], which introduce fewer privacy issues.

Earlier works on human activity recognition have leveraged signal processing techniques [8] and hand-engineered feature extraction methods [9]. Furthermore, traditional machine learning methods have also been widely adopted for human activity recognition in prior works. However, recent works have proposed various deep learning-based architectures that outperform the aforementioned works by extracting more complicated features from the input times-series data [2, 10, 11, 24, 13, 23, 22]. Considering the prior research on human activity recognition, we briefly summarize the involved challenges as follows:

1. **Deep Model Architecture Design**: There exists a wide range of complex deep learning architectures (such as feed-forward, convolutional [13], recurrent [14], residual [12], etc.). As each architecture has its own benefits and disadvantages, designing a model architecture that performs well for all human activity recognition datasets is challenging.

2. **Effective Time-Series Feature Extraction:** Prior works often consider time-series features to identify different activities. However, as shown in [2,10] spectral or statistical features could also serve as additional inputs to enhance the model's capabilities for more accurate human activity recognition. Therefore, there is a need to investigate the performance of different models given various types of features extracted from the sensor data to provide a clear understanding of their effectiveness.

3. **Efficient Model Training Mechanism:** Common human activity approaches rely on the traditional classification model training through the cross-entropy loss function. However, there exist other pretraining techniques including contrastive learning [15] or the triplet loss [16] that could further push the limits of the human activity model to generate better results.

In this work, we aim to perform extensive experimental studies on two human activity recognition datasets to measure the effectiveness of common deep learning architectures, feature extraction methods, and model learning techniques for human activity recognition. The rest of this paper is organized as follows. We first review the related work in Sec. II provide the details of the datasets and the preprocessing steps in Sec. III followed by the feature extraction and problem statement in Sec. IV. Then, we explore the studied model architectures and different learning mechanisms in Sec. V. We then present our experimental studies in Sec. V and conclude the paper in Sec. VI.

## II. RELATED WORK

Recent works on human activitiy recognition has been focusing on machine learning and deep neural networks due to their high accuracy for complicated tasks compared to hand-engineered works [8, 9]. For instance, the proposed method in [22], used the long short-term memory (LSTM) layers to extract the temporal information in the sensor time-series. Similarly, [21] added the attention mechanism on top of the LSTM layers to enhance the important feature extraction. Moreover, the model proposed in [13] is based on 1-dimensional convolutional neural network that extracts the temporal information from the sensor data in a more efficient way. The proposed model in [18] leverages two LSTM layers to process the time-series data in two direction to enhance the temporal information extraction of the model. The authors of [14] leveraged the residual connections to augment the training of the human activitiy recognition model. Furthermore, to improve the training quality of the model, the proposed method in [2] incorporates the contrastive learning loss function [15] in addition to the commonly used cross-entropy loss function, which ehnaces the representation learning of the model.

## III. DATASETS, DATA PREPROCESSING, AND FEATURE EXTRACTION

### A. Datasets

We briefly summarize the datasets studied as follows.

**Dataset 1** (DS1): The first dataset studied is collected by [17], which consists of 7,498 records and consists of six human activity classes as follows: going downstairs, walking upstairs, jogging, standing, walking, and sitting. This dataset contains the time-series data collected from the accelerometer sensor by 36 different users.

**Dataset 2** (DS2): The second dataset studied is provided by [11], which has a total of 39,168 records and consists of the following human activity classes: walking, bike riding, going upstairs, going downstairs, jogging, and riding in bux/taxi. This dataset contains the time-series data collected by 48 different users from both the accelerometer and gyroscope sensors.

### B. Data Preprocessing

Having the recorded time-series data of the accelerometer and gyroscope sensors along the x, y, and z axes, and with length $L$ = i.e., in $\mathbb{R}^L$, we perform the following pre-processing steps to prepare them for model processing.

**Segmentation:** Given the time-series data, we divide them into multiple segments with a sliding window of size $S$ (150 in this study), where each window has 70% of overlap with the previous window.

**Noise Filtering:** We use the moving average method to filter the noises caused by the vibrations that occurred while recording the sensor data. Specifically, we slide a window of size $M = 10$ and calculate the average of all the values within the window to eliminate the noise.

**Normalization:** Finally, since the scale of the values varies across different sensors, we leverage the min-max normalization to normalize each sensor axes (e.g., accelerometer along the y-axis) to have values in [0,1] interval.

## IV. FEATURE EXTRACTION AND PROBLEM STATEMENT

### A. Feature Extraction

We extract three different features from the time-series segments as described below:

**Temporal Features:** The most widely studied feature for human activity recognition is the temporal features [11, 2, 18]. Basically, each resultant segment after the pre-processing is considered as the temporal features, which are then commonly processed by recurrent neural networks to extract the temporal information within them. Since here we are focusing on the the two accelerometer and gyroscope sensors each producing values along three of the x, y, anx z axes, the temporal features for DS2 would have a dimension of $\mathbb{R}^{S \times 6}$, where $S$ is size of the time-series segment as stated earlier. The temporal features would have a size $\mathbb{R}^{S \times 3}$ for DS1 as it only contains the sensor data for the three axes of accelerometer.

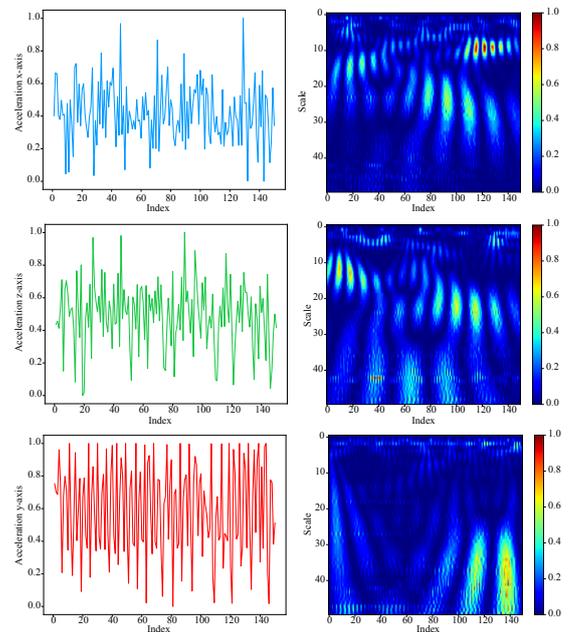

Figure. 1 Visualization of the temporal and spectral feature for one example from the jogging human activity class.

**Statistical Features:** Rather than processing the time-series segments, we can apply statistical functions (such as minimum, maximum, average, standard deviation, etc.) on each axis of each of the accelerometer and gyroscope sensors to extract statistical features. Such features have been shown by [10] to be highly effective for similar sensor time-series classification tasks. In this work, we consider the four minimum, maximum, average, and standard deviation functions to extract statistical features from each of the 6 axis of the accelerometer and gyroscope sensors. Thus, the resultant statistical features would have a dimension in $\mathbb{R}^{24}$ for DS2 and $\mathbb{R}^{12}$ for DS1.

**Spectral Features:** Finally, to capture more complicated patterns and extract far more advanced features, recent studies [2, 10], inspired by audio processing feature extraction methods [19], have proposed to extract spectral features from the time-series segments. Specifically, the continuous wavelet transform (CWT) function is applied on each sensor axis with different scales and the resultant features are combine to create multi-dimensional features. In this work, we leverage the Morlet wavelet function with 50 different scales to apply CWT on the time-series segment [20]. Therefore, the spectral features would have their dimension in $\mathbb{R}^{50\times S\times 6}$ for DS2 $\mathbb{R}^{50\times S\times 3}$ for DS1, where $S$ is the length of the segment as before.

To better demonstrate the extracted features, we visualize the temporal features and their correponding spectral features for the accelerometer time-series values along the x, y, and z axis of an example record belonging to the jogging human activity class in Fig. 1. We can observe that generally, the jogging exhibits a repeated pattern in the accelerometer time-series values due the nature of this activity.

*B. Problem Statement*

Given the sensor data recorded with the accelerometer and gyroscope sensor, the task of the human activity recognition model is to predict the correct human activity class. As stated above, in this work, the studied datasets consist of 6 different human activity classes.

V. MODEL ARCHITECTURES AND TRAINING MECHANISMS

In this section, we provide the details of the deep model architectures and the training mechanisms explored for the experimental studies.

*A. Model Architectures*

Considering the fact that the temporal and spectral features could be either be processed by recurrent or convolutional neural networks, we have adopted the various model architectures from the literature [2, 10, 11, 12 13, 14] to consider for our experiments. Besides, since traditional machine learning models are also widely studied for human activity recognition, we consider the common machine learning models for human activity recognition as baselines.

**Traditional Machine Learning Models:** We study support vector machine (SVM), K-nearest neighbor (KNN), gradient boosting decision tree (GBDT), logistic regression (LR), decision tree (DT), random forest (RF), AdaBoost, Gaussian Naïve Bayes (GaussianNB), and Multi-layer perceptron (MLP) as the most commonly used machine learning models for human activity recognition. The input to these models is the statistical features. For SVM, we use the linear kernel function. For KNN we set the number of the neighbors to 5. For RF we set the number of estimators to 100. Finally, for MLP we use two hidden layers.

**ResNet** [14]: We adapt the residual connection proposed in [24] to design a network based on convolutional neural neworks. Specifically, we use 4 residual blocks each having two convolutional and two residual layers. The input to this model is the spectral features.

**Transformers** [23]: Recently, transformers [23] have shown to be very effective in various domains. Thus, we have designed a neural network architecture based on the transformers that process the temporal features to identify different human activities. For this model, we leverage two transformer layers each having 8 heads.

**LSTM** [13, 22]: According to [13, 22], We design a network based on long-short term memory that processes the temporal features. We leverage one LSTM layer with 64 hidden units for this model.

**BiLSTM** [18]: To better capture the temporal information, we process them in both the forward and backward direction and leverage the combination of the features extracted from both directions to classify the human activities. We leverage 2 BiLSTM layers each having 64 hidden units for this model.

**LSTM-Attention** [21]: We augment the LSTM network previously stated with the attention mechanisms to measure the effects of such designs for human activity recognition. For this model, we leverage 2 LSTM layers with the attention mechanism where each has 64 hidden units

**CNN1D** [13]: Recurrent neural networks are often slow and involve high computation overheads. Thus, we have designed a network architecture based on 1-dimensional convolutional neural networks (CNN1D) to process the temporal features for human activity recognition. For this model, we stack two 1-dimensional convolutional layers with 64 and 32 filters, respectively. Besides, we set the kernel size of the convolutional layers to 3.

**MRNet** [2,10]: Inspired by prior studies, the combination of all the temporal, statistical, and spectral features could be effective for higher classification accuracy. Therefore, we have adapted the network proposed in [2,10] that first processes the temporal, statistical, and spectral features with sub-networks based on recurrent, fully connected, and convolutional neural networks. Then, we concatenate the output of all three networks and use them to predict the human activity class.

For all the models above, we use the rectified linear unit (ReLU) function as the activation function. Besides, we add three fully connected layers with 256, 128, and 6 hidden units as their last layer to perform human activity classification for a total of 6 different classes. Similarly, all the classification layers leverage the ReLU activation function while the last layer uses the softmax function to generate the probability values for each human activity class.

### B. Training Mechanisms

**Cross Entropy Classification Loss:** The most commonly used loss function for model training is the cross entropy loss formulated as follows:

$$\ell_{CE} = -\sum_{i=1}^{Z} p \log(\hat{p}), \quad (1)$$

where $p$ is the probability of the correct human activity class and $\hat{p}$ is the probability of the correct class generated by the model. $Z = 6$ is the total number of classes in this study.

**Supervised Contrastive Learning Loss:** Contrastive learning has been recently adopted for model pretraining in various tasks [15]. Here we adopt the supervised variant of the contrastive learning that leverages the label information from the dataset to generate distinguishable embeddings for each human activity class. The supervised contrastive learning loss is formulated as follows:

$$\ell_{CL} = \sum_{i \in Q} \frac{-1}{|A(i)|} \sum_{a \in A(i)} \log \frac{\exp(e_i \cdot e_p / \tau)}{\sum_{b \in A(i)} \exp(e_i \cdot e_b / \tau)}, \quad (2)$$

where $Q$ is the set of all the data records, $e_i$ is the embedding of the $i$-th data record, $\tau$ is the temperature parameter, $A(i)$ is the set of all the other data records with the same class as the $i$-th record.

**Triplet Loss:** Similarly, the triplet loss [16] aims to generate similar embeddings for data records belonging to the same class. The triplet loss is formulated as follows:

$$\ell_{TL} = \max(d(e_a, e_p) - d(e_a, e_n) + m, 0), \quad (3)$$

where $e_a, e_p, e_n$ are the embeddings of the anchor, positive (same class as the anchor), and negative (different class than the anchor), respectively, and $m$ is the margin controlling the distance between the embeddings. Besides, $d(\cdot)$ represents the distance function such as the Euclidean distance.

Based on the above, we train different models using the cross-entropy loss without pretraining. On the other hand, we can first pretrain the model based on either the contrastive or the triplet loss functions, and then continue the training based on the cross-entropy loss function.

### VI. Experimental Studies

In this section, we first review the parameter settings, and then present and discuss the experimental studies.

**Parameters:** For all the models, we use the Adam optimizer to train them with a learning rate of 0.001. We train the models using the cross-entropy for 50 iterations. For pretraining, we set the number of the iterations to 10. Besides, we set temperature parameter of the contrastive learning to $\tau = 0.07$. Moreover, we leverage 70% of the data for training, 10% for validation, and 20% for evaluation.

TABLE. 1 CROSS-ENTROPY ACCURACY (%)

| Model | DS1 | DS2 |
|---|---|---|
| SVM | 0.779 | 0.569 |
| KNN | 0.935 | 0.798 |
| GBDT | 0.892 | 0.784 |
| LR | 0.763 | 0.555 |
| DT | 0.874 | 0.759 |
| RF | 0.929 | 0.850 |
| AdaBoost | 0.446 | 0.683 |
| GaussianNB | 0.781 | 0.538 |
| MLP | 0.775 | 0.603 |
| ResNet | 0.954 | 0.535 |
| Transformers | 0.878 | 0.840 |
| LSTM | 0.953 | 0.873 |
| BiLSTM | 0.954 | 0.874 |
| LSTMAttention | 0.931 | 0.870 |
| CNN1D | 0.939 | 0.828 |
| MRNet | 0.970 | 0.552 |

TABLE. 2 SUPERVISED CONTRASTIVE LEARNING ACCURACY (%)

| Model | DS1 | DS2 |
|---|---|---|
| ResNet | 0.882 | 0.872 |
| Transformers | 0.852 | 0.813 |
| LSTM | 0.923 | 0.857 |
| BiLSTM | 0.915 | 0.856 |
| LSTMAttention | 0.870 | 0.826 |
| CNN1D | 0.919 | 0.820 |
| MRNet | 0.854 | 0.723 |

**Performance Results:** We first train the model based on the cross-entropy loss function on DS1 and DS2 and illustrate the results in Table. 1 We can see that the high accuracy on DS1 is achieved by the ResNet and LSTM model. Although the performance achieved by these two models is very close, we realized that since ResNet takes advantage of the residual connections and is based on lightweight convolutional neural networks, compared to LSTM, the training procedure was much shorter as the model converged faster. Furthermore, we can observe that MRNet has outperformed both the ResNet and LSTM models by combining the temporal, spectral, and statistical features, and further supports the idea proposed in [2, 10].

On the other hand, we can see that the recurrent neural models such as LSTM and BiLSTM have achieved much higher performance on DS2 compared to the other models, which shows that such models are more suitable for learning on large datasets. Besides, we can observe that the traditional machine learning models such as SVM, AdaBoost or GaussianNB have generally achieved low performance due to their limited computation capabilities.

TABLE. 2 TRIPLET LOSS ACCURACY (%)

| Model | DS1 | DS2 |
| --- | --- | --- |
| ResNet | 0.779 | 0.569 |
| Transformers | 0.935 | 0.798 |
| LSTM | 0.892 | 0.784 |
| BiLSTM | 0.763 | 0.555 |
| LSTMAttention | 0.874 | 0.759 |
| CNN1D | 0.929 | 0.850 |
| MRNet | 0.954 | 0.535 |

Next, we perform the experiments by pretraining the models based on the supervised contrastive learning and the triplet loss and show the results in Tables. 2 and 3, respectively. We can observe no performance improvements on DS1 and an improvement less than 1% on DS2 for some of the models based on these pretraining methods.

In summary, the BiLSTM model with the temporal features as its input and the cross-entropy loss function has the highest accuracy among all the deep learning models. We illustrate the confusion matrices of the BiLSTM model for DS1 and DS2 in Fig. 2, which shows the high accuracy of this model for different human activity classes.

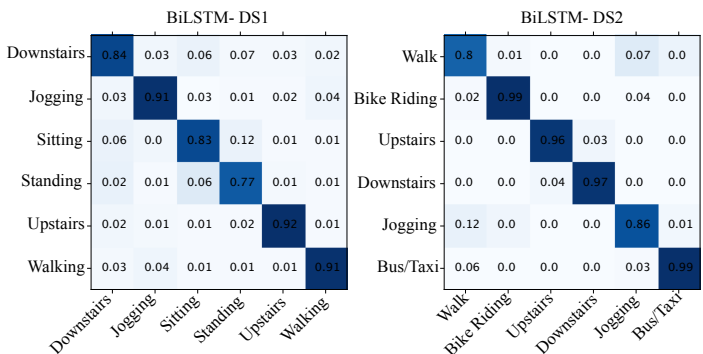

Figure. 2 The confusion matrices of the most accurate model (BiLSTM) on DS1 and DS2.

## VII. Conclusion

In this paper, we investigated the performance of different deep neural network architectures for human activity recognition given the temporal, statistical, and spectral features. Moreover, we explored different model designs based on residual connections, convolutional and recurrent layers, transformers, attention mechanisms, and traditional machine learning algorithms. Moreover, we trained the models based on three common learning algorithms and compared their performance according to our experiments on two large-scale human activity recognition datasets. According to our results, the combination of multiple features could lead to performance improvements while learning algorithms such as contrastive learning or the triplet loss could be less effective depending on the complications within the dataset.